\documentclass{ws-p8-50x6-00}

\def\.{\cdot}

\def\o{\over}

\def\d{\delta}

\def\u{\omega}

\def\o{\over}
\def\p{\partial}

\def\+{\bigoplus}

\def\({\left(}
\def\){\right)}
\def\[{\left[}
\def\]{\right]}
\def\l.{\left.}
\def\r.{\right.}

\def\be{\begin{equation}}
\def\ee{\end{equation}}
\def\bea{\begin{eqnarray}&&}
\def\eea{\end{eqnarray}}
\def\nn{\nonumber \\ &&}

\def\ber{\begin{array}}
\def\eer{\end{array}}

\begin{document}

\title{The Functional Measure of Gauge Theories in the Presence of Gribov
Horizons\footnote{Talk given by C. Becchi at the conference ``Path Integrals from peV to TeV'', Florence-Italy, August 25-29, 1998}}

\author{Carlo Becchi, Stefano Giusto}

\address{ Dipartimento di Fisica dell' Universit\`a di Genova,
Via Dodecaneso 33, I-16146, Genova, Italy}

\author{Camillo Imbimbo}

\address{ INFN, Sezione di Genova,
Via Dodecaneso 33, I-16146, Genova, Italy}  


\maketitle

\abstracts{We discuss the formal structure of a  functional measure for Gauge Theories
preserving the Slavnov-Taylor identity in the presence of Gribov
horizons. Our construction defines a gauge-fixed measure in the
framework of the lattice regularization by dividing the configuration space
into patches
with different gauge-fixing prescriptions. Taking into account the bounds
described by
Dell'Antonio and Zwanziger we discuss the behaviour of the measure in the
continuum limit for
finite space-time volume.
}

The purpose of this note is to define a suitable framework to study the
continuum-weak-coupling limit of lattice gauge theories; that is the limit
in which locally the gauge fields fluctuations are small in comparison with
the inverse product of the bare coupling $g_0$ times the lattice spacing.
In this limit the non-perturbative Wilson lattice theory is believed
to approach in some sense the short-distance covariant perturbation theory.

The considered limit is formally obtained from Wilson's theory making the
lattice spacing to vanish with respect to a given scale $\mu^{-1}$ and
keeping the product $g_0^2\log{1\over a\mu}$ constant. To avoid the 
overwhelming
difficulties connected with the infra-red limit we keep the space-time
volume of the
lattice  constant and smaller than $\mu^{-D}$ in $D$ space-time dimensions.
The topological structure of the configuration space is reasonably
well known  at
finite volume \cite{1},  while it is expected to develop  catastrophic
singularities in the
infinite volume limit.

To limit the field fluctuations that are longitudinal with respect to the
gauge orbits we introduce a suitable gauge-fixing condition. Only those 
conditions that are complete and
translational invariant induce a sufficient suppression of the longitudinal
field fluctuations. For example the temporal gauge: $A_0=0$ is not suitable
for our purposes.

Our strategy is based on the introduction of gauge-fixing terms, together
with the corresponding Faddeev-Popov determinants, already in the lattice
functional measure in order to have a uniform limitation of the field
fluctuations in the continuum limit.  However in this limit the situation
is made cumbersome by the appearance of Gribov horizons \cite{2}; the
Faddeev-Popov
determinant  is expected to  become degenerate
 for a non-trivial subset of configurations. In this case the 
gauge-fixing procedure
would encounter a singularity and hence the functional measure would appear
ill-defined. It
is for this reason that it has been speculated \cite{3} that the Gribov
phenomenon could
induce a breakdown of the Slavnov-Taylor identity that ensures the
consistency of the
Faddeev-Popov quantization procedure. In perturbation theory this breakdown
would not appear
since also the Gribov horizons are absent.

We want to show, in contrast with these speculations, that there exists a
lattice gauge-fixed functional measure based on a decomposition of the
configuration space into patches equipped with different gauge-fixing
prescriptions for which the Slavnov-Taylor
identity is satisfied.  For a suitable choice of the patch decomposition
this measure
should  avoid the  Gribov horizons \cite{4}.

Requiring the functional measure to obey the Slavnov-Taylor identity
ensures two main results. First, whenever a theory leads to a space of 
asymptotic spaces this identity guarantees the unitarity of the
restriction of the $S$-matrix to a suitable physical subspace \cite{5}; more
generally it ensures the invariance of the correlators of physical
operators under deformations of the  gauge-fixing prescriptions \cite{6}.
This implies in particular the independence of physical
correlators on the choice of the patch decomposition of the
configuration space. Notice that this does not
exclude the existence of physical effects induced by the  very presence of
Gribov horizons; indeed the atlas of patches can be modified but not
reduced to a single patch, at least in the continuum limit.

For simplicity we choose in our study the Landau background gauge
prescription.
Given two gauge configurations $\{U_l\}$ and
$\{V_l\}$, where $l$ labels the lattice links,
we define the relative distance as:
\be
d^{2}\(U,V\)=\sum_{l}\mathrm{Tr}\[\( U_{l}-V_{l}\) \(
U_{l}^{\dag}-V_{l}^{\dag}\)\].\label{dl}
\ee
This defines the lattice $L^{2}$ norm which tends
in the continuum limit to the norm:
\be
d^{2}\(A,B\)=\int dx {\mathrm Tr}\[\( A^{\mu}-B^{\mu}\) \(
A_{\mu}^{\dag}-B_{\mu}^{\dag}\)\] \ ,\label{dc}\ee
where $A$ and $B$ are the matrix valued gauge fields corresponding to
$U$ and $V$ respectively. The Landau gauge-fixing prescription associates
every gauge orbit with the configuration on the orbit for which the distance
from the background has an absolute minimum. The neighborhood of the background
for which this minimum is unique is called its "fundamental modular region".
We associate a sphere with radius $R$ in the norm defined above
to each background gauge configuration.
The radius $R$ of the sphere corresponding to every background is such
that in the continuum limit the sphere does not intersect the horizon
of the Gribov domain corresponding to the Landau choice.  Indeed it is
known that in the continuum limit and at finite space-time volume the
fundamental modular region contains a sphere whose radius is
independent of the background \cite{7}.

It remains to choose a suitable set of background configurations such
that every orbit intersects at least one sphere of radius $R$ centered
in an element of the set. This can be done on the lattice owing to
the existence of global gauge-fixings.  Indeed one can choose, for
example, the configurations for which $U_l$ is constant along every
lattice link parallel to the first coordinate axis and also
constant along those links parallel to the second axis which belong
to a given hyperplane orthogonal to the first axis.
Furthermore $U_l$ can be taken constant
along the links parallel to the third coordinate axis and belonging to
a plane orthogonal to the first two axis. Finally the $U_l$ are
chosen constant along a single lattice line parallel to the fourth axis.
This is a complete axial gauge prescription that, as mentioned above,
does not induce the
wanted limitation of the longitudinal field fluctuations in the
continuum limit thus being unsuited for studying the continuum
limit. However we can use this gauge-fixing to select the wanted background
configurations.  Indeed since these gauge-fixed configurations define
a compact manifold $M_{L}$, we can choose a finite lattice of
configurations on this manifold such that the distance between two
neighboring elements is smaller than $\sqrt{2} R$ and such that the
corresponding set of spherical patches of radius $R$ centered in the
elements of the configuration lattice covers $M_{L}$ completely.

We now come to a formal construction of the lattice measure.  We
consider a periodic hypercubic space-time lattice with lattice spacing
$a$ and period $La$.  We label by $x$ a generic lattice site whose
coordinates are: $x^{\mu}=a n^{\mu}$, where $n$ is a four-vector with
integer components.  Let $e_{\mu}$ be the vector with components:
$e_{\mu}^{\nu}=a \d_{\mu}^{\nu}$. Then $x+e_{\mu}$ is the nearest
neighboring site to $x$ in the direction of the $\mu$ axis.  We
limit our study to the gauge group $SU(N)$.  A lattice configuration corresponds
to the set of unitary, unimodular, $N$-dimensional matrices:
$\{U_{x,\mu}\}$ where the index $x,\mu$ labels the link joining the sites
$x$ and $x+e_{\mu}$.  A gauge transformation is given by the set of
unitary, unimodular, $N$-dimensional matrices $\{g_{x}\}$ satisfying
the constraint $g_{o}=I$, where $o$ labels the space-time origin and $I$
is the unity matrix. This constraint distinguishes the gauge
transformations from the $SU(N)$ rigid symmetry of the model.
The action of a gauge transformation on the configuration $U$ is given
by:
\be
U\longrightarrow U^{(g)}\quad\quad , \quad  U^{(g)}_{x,\mu}=g_{x}
U_{x,\mu}g^{\dag}_{x+e_{\mu}}\ .
\ee
Given a background configuration $V$, the gauge-fixed configuration
$U^{(\bar g)}$ corresponding to the orbit
$U^{(g)}$ is the one for which $d^{2}\(U^{(g)}, V\)$ has an absolute minimum.

In general only a subset of the orbits admits a unique gauge-fixed
configuration corresponding to a given background. Indeed it is
possible that the same orbit contains two or more configurations
minimizing the distance, or that $V$ coincides with the center of an
osculator circle to the orbit. However it is clear that the
minimizing configuration is unique for the orbit  $V^{(g)}$ itself,
and thus is also unique for all the orbits crossing a small ball around $V$.
The work of Dell'Antonio and Zwanziger \cite{7} implies 
that the radius of this ball
can be chosen to be
independent of $V$ and of the number of lattice points for a given 
space-time volume.

An isolated minimizing configuration satisfies the local gauge condition:
\bea
\sum_{\mu}\(V^{\dag}_{x+e_{\mu},\mu}U_{x+e_{\mu},\mu}-U^{\dag}_{x+e_{\mu} ,
\mu}V_{x+e_{\mu} , \mu}\) \nn
 = \sum_{\mu}\(U_{x , \mu}V^{\dag}_{x , \mu}-V_{x+e_{\mu} ,
\mu}U^{\dag}_{x+e_{\mu} , \mu}\)\
,\label{gcl}\eea 
that in the continuum becomes:
\be {\cal D}^{(V)}_{\mu}\[A^{\mu}-V^{\mu}\]=0\ ,\label{gcc}\ee 
where ${\cal D}^{(V)}$ is the covariant derivative in the background $V$. This
is the well known Landau background gauge condition. (\ref{gcl}) and
(\ref{gcc}) define local sections $\Sigma_{V}$ of the gauge bundle
whose total space is the configuration space.


\begin{figure}[t]
\epsfxsize=15pc 
\begin{center}
\epsfbox{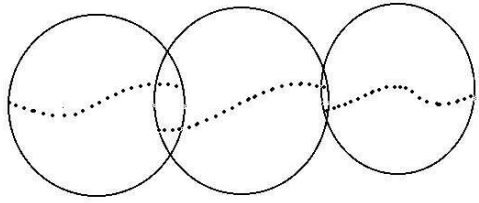} 
\caption{The circles represent the patches and the dotted lines
the local sections.}
\end{center}
\end{figure}

We select, as described above, a collection of background
configurations $\{V^{(a)}\}$
that is finite on the lattice, but becomes infinite in the continuum
limit. We also define the characteristic functions of the corresponding
patches in the following way. We call ``Gauge Strip''
the union of the spherical patches around the
background configurations  $V^{(a)}$; its characteristic function is:
\be
\chi_{\mathrm{strip}}(U)\equiv
\Theta\( R-\mathrm{inf}_{a}d(U, V^{(a)})\),\label{strip}
\ee
where $\Theta$ is a smoothened Heavyside function. Then,
the characteristic function of the patch around $V^{(a)}$ is:
\be
\chi_{a}(U)\equiv \chi_{\mathrm{strip}}(U){ \Theta\( R-d\(U, V_{a}\)\)\o
\sum_{b}\Theta\( R-d\(U, V_{b}\)\)}.\label{pa}
\ee
Thus
\be\sum_{a}\chi_{a}(U)=\chi_{\mathrm{strip}}(U).
\ee

We assume without proof that {\it the  configurations in the intersection
of the  Landau local sections
$\Sigma_{a}$ with the corresponding patches are internal to the Gauge
Strip.}  The meaning of this assumption is illustrated in the figure.
\noindent

The proof of the above assumption can be given following the study of the
covariant
Laplacian presented in \cite{7}.

To construct the wanted functional measure we introduce the exterior (BRS)
derivative $s$:
\bea sU_{x , \mu}=i\(\u_x U_{x , \mu}-U_{x , \mu}\u_{x+e_\mu}\)\nn
s\u_x=i\u_x^2\quad\quad\quad\quad\quad s\bar\u_x=i b_x\nn
sV_{x , \mu}=0\ .\label{brs}\eea
where $b_x$ is the Lagrange-Nakanishi-Lautrup multiplier implementing the 
gauge-fixing constraint.
We also introduce the anti-derivative:
\bea \bar sU_{x , \mu}=i\(\bar\u_x U_{x , \mu}-U_{x , \mu}\bar\u_{x+e_\mu}\)\nn
\bar s V_{x , \mu}=0\ .\label{anti}\eea
The standard Faddeev-Popov measure corresponding to the background $V_a$ is:
\be d\mu_{\mathrm inv}\prod_x d\u_x d\bar\u_x d b_x e^{-s\bar s
d^2\(U,V\)}\equiv d\mu
e^{-s\Psi_a}\ ,\label{mfp}\ee
where $d\mu_{\mathrm inv}$ is the lattice gauge invariant measure and we have
identified the
gauge-fixing fermionic operator $\Psi_a$ with ${\bar s} d^2\(U,V\)$.

Naively, the measure corresponding to the chosen patch decomposition of the
configuration space would be:
\be
d\mu_{\mathrm naive}=\sum_a\chi_a d\mu e^{-s\Psi_a}\, .
\ee
This measure, however, does not satisfy the  Slavnov-Taylor identity,
stating that
\be \int d\mu_{\mathrm naive} s\Omega=0 \label{st}
\ee
for any field functional $\Omega$, which is required for the physical
reasons we discussed above.

One has instead:
\be \int d\mu_{\mathrm naive} s\Omega=-\sum_a\int d\mu s\chi_a
e^{-s\Psi_a}\Omega\ ,\ee
whose right-hand side does not vanish since the  characteristic functions of the
patches are not gauge invariant.

This difficulty can be overcome by adding to the naive measure
contributions associated to the intersections of different patches.
Consider for simplicity the case in which one has only two patches.
On account of our assumption, on the support of the measure $d\mu
e^{-s\Psi_a}$ one has:
$\chi_1+\chi_2 = 1$, which implies
\be s\chi_1=-s\chi_2=s\chi_1\chi_2-s\chi_2\chi_1\ .\ee
Therefore one has:
\bea \int d\mu_{\mathrm naive}\, s\Omega=-\int d\mu s\chi_1\( e^{-s\Psi_1}-
e^{-s\Psi_2}\)\Omega\nn=\int d\mu \(s\chi_1\chi_2-s\chi_2\chi_1\)
s\(\Psi_1-\Psi_2
\) \int_0^1 dt\  e^{-s\(t\Psi_1+(1-t)\Psi_2\)}\Omega\nn=
\int d\mu \(s\chi_1\chi_2-s\chi_2\chi_1\) \(\Psi_1-\Psi_2
\) \int_0^1 dt\  e^{-s\(t\Psi_1+(1-t)\Psi_2\)}s\Omega\ .\eea
We thus see that in the case of two patches the measure:
\bea
d\mu \[ \chi_1 e^{-s\Psi_1} + \chi_2
e^{-s\Psi_2}\r.\nn\l.-\(s\chi_1\chi_2-s\chi_2\chi_1\) \(\Psi_1-\Psi_2 \)
\int_0^1 dt\  e^{-s\(t\Psi_1+(1-t)\Psi_2\)}\]\ .
\eea satisfies the Slavnov-Taylor identity.
It  is clear that the term added to the
naive measure is supported in the intersection of the two patches. This
result can be extended to the case of an atlas with a generic number of
patches. To this end, we define $n$-chart interpolating measure:
\be
e^{-s\Psi_{a_{1},...,a_{n}}}\equiv\int_0^\infty\prod_{i=1}^ndt_i\
\d\bigl(\sum_{j=1}^nt_j-1\bigr)
e^{-s\sum_{k=1}^n t_k \Psi_{a_k}}\ .\label{def1}
\ee
and we introduce the two following notations:
\be \bigl(s\chi_{a_{1}}...s\chi_{a_{n-1}}\ \chi_{a_{n}}\bigr)_A\equiv
\sum_{k=1}^n(-1)^{k-n}\chi_{a_{k}}s\chi_{a_{1}}...s\check\chi_{a_k}...s\chi_
{a_{n}}\,\label{def2}
\ee
and:
\be \p_{\Psi}\(\Psi_{a_{1}}...\Psi_{a_{n}}\)\equiv\sum_{l=1}^n(-1)^{l+1}
\Psi_{a_{1}}...\check\Psi_{a_l}...\Psi_{a_{n}}\ .\label{def3}\ee
In (\ref{def2}) and (\ref{def3}) the check mark above  a factor means
that the corresponding factor should be omitted.
Then the functional measure consistent with the Slavnov-Taylor
identity is:
\be
d \mu_{\mathrm ST}=d\mu\sum_{n=1}^\infty {(-)^{(n-2)(n-1)\o2}\o n}
\(s\chi_{a_{1}}...s\chi_{a_{n}}\ \chi_{a_{n}}\)_A
\p_{\Psi}\(\Psi_{a_{1}}...\Psi_{a_{n}}\)e^{-s\Psi_{a_{1},...,a_{n}}}
\label{mes}
\ee
This is the main result presented in this communication.  It is apparent
that the lack of gauge invariance of the characteristic functions of the
patches induces new contributions to the measure localized on the patch
(regularized) boundaries. One can interpret the
factors multiplying the interpolating measure in the patch intersections as
the Jacobians relating the bulk measure to the boundary one. Before the
continuum limit, the number of patches being finite, the series appearing  in
(\ref{mes}) reduces to a finite sum. However in the continuum limit
 our patch decomposition could  present intersections
of an infinite number of patches.  In this case the meaning of our
formula would be unclear since we should have to face convergence
problem for our series. As a matter of fact, if the space-time volume
remains finite, the configuration space in the continuum loses its
compactness but remains paracompact even in the $L^2$ norm
\cite{8}. This means that one can avoid these convergence
problems at finite space-time volume.

Of course, in the infinite volume limit, the above measure could be
ill defined if the cells  accumulated around some singularity of
the configuration space. This could perhaps induce instabilities of
the BRS symmetry in the sense of \cite{3}. We believe however that,
in analogy with the topological case, these instabilities could be
reabsorbed extending the action of the BRS operator on the moduli
space of the infra-red singular configurations.

Its is easy to verify directly that the expectation values of ``physical
functionals",
that are annihilated by $s$, do not vary if the patch decomposition is deformed
continuously, provided one avoids Gribov horizons; this remark substantiates the
above consideration on the physical consequences of the Gribov phenomenon.
It would be
interesting to verify which is the relevance of the  patch intersections
closest to the
trivial vacuum configuration to a physically significant expectation value.

\subsection*{Aknowledgements} CB wishes to thank M.Asorey and R. Stora for
interesting remarks.

\end{document}